\documentclass[prd,showpacs,nofootinbib,preprintnumbers,amsmath,amssymb]{revtex4}

\usepackage{graphicx}
\usepackage{dcolumn}
\usepackage{bm}

\begin{document}

\title{QED Actions in Supercritical Fields}

\author{Sang Pyo Kim}\email{sangkim@kunsan.ac.kr}
\affiliation{Department of Physics, Kunsan National University, Kunsan 573-701, Korea}
\affiliation{Center for Relativistic Laser Science, Institute for Basic Science (IBS), Gwangju 500-712, Korea}

\author{Hyun Kyu Lee}\email{hyunkyu@hanyang.ac.kr}
\affiliation{Department of Physics, Hanyang University, Seoul 133-791, Korea}

\medskip

\date{\today}

\begin{abstract}
In the in-out formalism we advance a new method to represent the gamma function for QED actions in supercritical fields, which is complementary to the proper-time integral representation in Phys. Rev. D {\bf 78}, 105013 (2008) and Phys. Rev. D {\bf 84}, 065004 (2011). The new method directly yields the QED action in terms of the Hurwitz zeta function in a constant magnetic field and the complex QED action in a constant electric field. The complex action exactly gives the vacuum polarization and the vacuum persistence and thereby the pair-production rate in the electric field. The QED actions exhibit the electromagnetic duality.
\end{abstract}
\pacs{11.15.Tk, 12.20.Ds,  13.40.-f}

\maketitle

\section{Introduction}

In quantum electrodynamics (QED) a strong electromagnetic field is known to polarize the Dirac vacuum and create charged particle-antiparticle pairs. To understand the quantum vacuum influenced by an external field, one employs the
effective action obtained by integrating out the field of a charged particle, which depends basically on the mass and charge of the particle in addition to the field. One prominent feature of QED action is the emergence of the imaginary part (vacuum persistence) in the presence of an electric field \cite{heisenbergeuler36,weisskopf36,schwinger51-1}, a consequence of Schwinger pair production. In fact, the Dirac vacuum in a constant electric field becomes unstable against pair production when virtual pairs can be effectively separated over one Compton wavelength at the cost of the electric potential energy for each pair. Schwinger pair production thus prohibits stable electric fields from being attained through physical processes beyond the critical limit, $E_{\rm C} = m^2/q, $\footnote{The natural unit of $\hbar = c =1$ is employed.} the lowest limit being $ m_e^2/e ~( 1.3 \times 10^{16}~{\rm V/cm})$ for electrons among the known charged particles in the standard model.

However, physical processes do not prevent magnetic fields from being accumulated beyond the critical strength $B_{\rm C} = m_e^2/e ~( 4.4 \times 10^{13}~{\rm G})$, but magnetic fields would not increase beyond another critical value $B_{\rm M} = \alpha (m_m/m_e)^2 B_{\rm C}$ for monopole production of mass $m_m$ \cite{affleckmanton82}. Magnetars, a class of neutron stars, are believed to have magnetic fields of $10^{14}-10^{16}~{\rm G}$ on the surface and even higher intensity in the core \cite{hardinglai06}, and strong QED phenomena have been studied in compact stars \cite{rvx}.
In condensed matter physics, the effective theory for electrons and holes in an applied electromagnetic field has an analogous critical value, which is determined by the effective mass, much smaller than the rest mass. The QED analogous phenomena thus become important in graphene \cite{graphene05,graphene09} as well as dielectric materials \cite{okaaoki09}. Hence, QED actions in supercritical magnetic fields are not only a theoretical interest but also an issue of physical applications, and computing QED actions in supercritical electric fields is a theoretically intriguing issue due to the complex actions.

Computing the QED action in any electromagnetic field has been a challenging task since the seminal works on a uniform electromagnetic field by Heisenberg-Euler, Weisskopf and Schwinger \cite{heisenbergeuler36,weisskopf36,schwinger51-1}. Various methods have been developed to compute the determinants of the Dirac or Klein-Gordon operators in electromagnetic fields but applied only to a few configurations (see Refs. \cite{gmr,dittrichreuter,dittrichgies,dunne05}). Recently, the in-out formalism based on the variational principle by Schwinger and DeWitt \cite{schwinger51-2,dewitt75,dewitt03} has been elaborated to find the QED actions in the uniform electromagnetic field or Sauter-type electric or magnetic fields \cite{kly08,kly10,kim11}. The scattering matrix between the out-vacuum and the in-vacuum is expressed in terms of gamma functions with complex arguments and leads to the proper-time integral \cite{kly08,kly10,kim11} or the power series \cite{ahn83,nikishov03} for QED action.

The QED Lagrangian density, simply action, in a constant magnetic or electric field is given by the proper-time integral \cite{dunne05}
\begin{eqnarray}
{\cal L}^{(1)} (\kappa) = - \Bigl( \frac{m^2}{4 \pi \kappa} \Bigr)^2 \int_{0}^{\infty}
\frac{ds}{s^2} e^{-\kappa s } F_{\kappa} (s), \label{prop in}
\end{eqnarray}
where $\kappa$ denotes the inverse of dimensionless field strength
\begin{eqnarray}
\beta = \frac{m^2}{2qB} = \frac{B_{\rm C}}{2 B}, \quad \epsilon = \frac{m^2}{2qE} = \frac{E_{\rm C}}{2 E}, \label{para}
\end{eqnarray}
and the spectral function for the magnetic field is
\begin{eqnarray}
F_{\beta} (s) = \frac{\cosh(s/2)}{\sinh (s/2)}
- \frac{2}{s} - \frac{s}{6},
\end{eqnarray}
and that for the electric field $F_{\epsilon} (s) = i F_{\beta} (is)$. The one-loop action (\ref{prop in}) as a Laplace transformation of $F_{\kappa} (s)/s^2$ may be used for a numerical purpose even for supercritical fields $(\beta,~ \epsilon \leq 1)$. However, when the spectral function is expanded in a power series \cite{gr-table}, QED action does not converge for supercritical fields because the radius of convergence decreases to zero for higher terms, in other words, QED action is asymptotically divergent \cite{dunne05}. Furthermore, the exponential function in Eq. (\ref{prop in}) plays the role of a regulator for the proper integral, so the supercritical field determines the behavior of the upper limit of the integral. To have useful expressions for supercritical fields, several methods have been employed to represent QED actions in terms of the Hurwitz zeta function \cite{dittrich76,dittrich77,bvw91}, the gamma function \cite{dtz79,vlm82,heylhernquist97}, and a convergent series in constant subcritical electromagnetic fields \cite{mielniczuk82,vlm93,vlm94,chopak01,bcp01,jgvkw01,ksx13}.

In this paper we advance an entirely new method to compute the QED action for a constant supercritical magnetic or electric field. For that purpose, we employ the in-out formalism and express the one-loop effective action in terms of the gamma function. In the previous works the gamma function has provided a representation of the proper-time integral \cite{kly08,kly10,kim11} and also a power series \cite{ahn83,nikishov03}, equivalent to the QED action (\ref{prop in}). Now we directly use the Ramanujan formula for the gamma function \cite{bernadt} to express the QED action in a power series and show the equivalence with the power series of Eq. (\ref{prop in}). We further apply the Whittaker-Watson formula \cite{whittakerwatson} to obtain the QED actions for the supercritical magnetic or electric field in terms of the Hurwitz zeta function. Remarkably, our method yields the QED action without summing over Landau levels and introducing a cutoff parameter in the magnetic field and directly recovers the complex QED action in the electric field. The electromagnetic duality and the consistency between the vacuum persistence and the pair-production rate in the electric field
follow as a consequence of the QED actions. The fact that all the representations of the gamma function have the same renormalization terms may shed light on the renormalization issue related to QED actions.

The organization of this paper is as follows. In Sec. \ref{sec2}, using the gamma function representations,
we find the QED action in a constant magnetic field in two different expressions: the power series (\ref{sp ser}) and the Hurwitz zeta function (\ref{zeta B-act}). In Sec. \ref{sec3}, we directly obtain the complex QED action (\ref{zeta E-act}) in a constant electric field in terms the Hurwitz zeta function. In Sec. \ref{sec4}, we discuss the physical implications and summarize the results.

\section{Constant Magnetic Field}\label{sec2}

In the in-out formalism by Schwinger and DeWitt \cite{schwinger51-2,dewitt75,dewitt03}, the scattering matrix ${\cal S} = \langle {\rm out} \vert {\rm in} \rangle$ gives the exact one-loop action
\begin{eqnarray}
\int d^3 x d t {\cal L}^{(1)} = -i \sum_{\bf K} \ln (\mu_{\bf K}^*), \label{in-out ac}
\end{eqnarray}
where $\mu_{\bf K}$ is the coefficient of the Bogoliubov transformation, $\hat{a}_{{\rm out} K} = \mu_{\bf K} \hat{a}_{{\rm in} K} + \nu_{K}^* \hat{b}^{\dagger}_{{\rm in} K}$, and ${\bf K}$ stands for all quantum numbers, such as the energy-momentum and spin states of the field. In a constant magnetic field the unrenormalized QED action takes the form \cite{kim11}
\begin{eqnarray}
{\cal L}^{(1)} (\beta) = - \frac{m^2}{(2 \pi)(2 \beta)} \sum_{\sigma} \int \frac{d \tilde{\omega}}{(2 \pi)} \frac{dk_z}{(2 \pi)} \ln (\Gamma (p^*)).
\label{qed ac}
\end{eqnarray}
where $\sigma= \pm 1/2$ are the spin states, $\tilde{\omega} = -i \omega$ is the Wick-rotated energy, and
\begin{eqnarray}
p = \frac{1 -2 \sigma}{2} + \frac{m^2 + \tilde{\omega}^2 + k_z^2}{(m^2/\beta)}.
\end{eqnarray}
In Ref. \cite{kim11}, an integral representation of the gamma function is used to express the QED action in the proper-time integral. We now employ other two representations of the gamma function in the QED action (\ref{qed ac}) and find the QED action in a power series and then in the Hurwitz zeta function below.

First, by using the Ramanuja formula in terms of the Bernoulli numbers $B_{k}$ \cite{bernadt}
\begin{eqnarray}
\ln \Gamma (z+1) = \Bigl(z  + \frac{1}{2} \Bigr) \ln z - z + \frac{1}{2} \ln (2\pi)
+ \sum_{k=1}^{\infty} \frac{ B_{2k}}{(2k)(2k-1) z^{2k-1}}, \quad (|{\rm arg} z| < \pi)
\label{log gam1}
\end{eqnarray}
and performing the integral over the momentum and energy, we obtain the QED action
\begin{eqnarray}
{\cal L}^{(1)} (\beta)= \frac{m^4}{(2 \pi)^2 (2 \beta)} \sum_{k =2}^{\infty} \frac{B_{2k}}{(2k)(2k-1)(2k-2)} \Bigl(\frac{1}{\beta} \Bigr)^{2k-1}.
\label{sp ser}
\end{eqnarray}
Note that the representation (\ref{sp ser}), a series expansion of the logarithm of gamma function (\ref{log gam1}), is equivalent to the proper-time representation (11) of Ref. \cite{kim11}, an integral formulation of the logarithm of gamma function. In the above, Schwinger's subtraction scheme is employed for renormalization of the action (\ref{qed ac}) and, for instance, the energy-momentum integral for spinor QED
\begin{eqnarray}
{\cal L}^{(1)}_{\rm div} (\beta)= \frac{m^4}{(2 \pi)^2 (2 \beta^2)} \int_{0}^{\infty} du \Bigl[ (u + \beta) \ln (u+ \beta) - (1+i \frac{\pi}{2}) (u+ \beta) + \frac{B_2}{2(u+ \beta)} \Bigr]
\label{ren}
\end{eqnarray}
has a UV-divergence and should be regulated away through the renormalization of the vacuum energy and the charge.
However, the series (\ref{sp ser}) is asymptotically divergent \cite{dunne05} and thus one needs other formula for the gamma function in the region of supercritical magnetic fields $\beta \leq 1$, which is the main motivation of this paper.

Second, using the Whittaker-Watson formula \cite{whittakerwatson}
\begin{eqnarray}
\ln \Gamma (z) = \Bigl(z- \frac{1}{2} \Bigr) \ln z - z + \frac{1}{2} \ln (2 \pi)
+ \frac{1}{2} \sum_{k = 1}^{\infty}
\sum_{n=1}^{\infty} \frac{k}{(k+1)(k+2)} \frac{1}{(z+n)^{k+1}}, \label{log gam2}
\end{eqnarray}
and $\ln \Gamma (z+1) = \ln \Gamma (z) + \ln z$, the renormalized action is given in terms of the Hurwitz zeta functions
\begin{eqnarray}
{\cal L}^{(1)} (\beta)=  \Bigl( \frac{m^2}{4 \pi \beta} \Bigr)^2 \Biggl[ \sum_{k =2}^{\infty} \frac{\zeta (k, 1+ \beta)}{(k+1)(k+2)}
+ \frac{1}{6} (\ln (\beta) - \psi(1+ \beta)) \Biggr].
\label{sp con}
\end{eqnarray}
Here, $\psi(z) = (d\Gamma(z)/dz)/\Gamma(z)$ is the psi-function, and the last terms in Eqs. (\ref{sp con}) originate from $\zeta(1, z)$, which renormalizes the charge in Eq. (\ref{ren}).
Finally, by summing over the Hurwitz zeta functions \cite{ccs04,zeta-der}, we obtain the renormalized action for spinor QED
\begin{eqnarray}
{\cal L}^{(1)} (\beta)=  \Bigl( \frac{m^2}{4 \pi \beta} \Bigr)^2 \Bigl(2 \zeta'(-1, \beta) -
(\beta^2 - \beta + \frac{1}{6}) \ln (\beta) + \frac{\beta^2}{2} - \frac{1}{6} \Bigr).
\label{zeta B-act}
\end{eqnarray}
The action (\ref{zeta B-act}) is consistent with the result from the zeta function regularization \cite{dittrichreuter,dunne05}
and recovers Eq. (\ref{sp ser}) by expanding the derivative of the Hurwitz zeta function in the weak-field limit.

\section{Electric Field and Electromagnetic Duality} \label{sec3}

The unrenormalized QED action in a constant electric field is given by \cite{kly08,kly10}
\begin{eqnarray}
{\cal L}^{(1)} (\epsilon) = i \frac{m^2}{(2 \pi)(2 \epsilon)} \sum_{\sigma} \int \frac{d^2 {\bf k}_{\perp}}{(2 \pi)^2} \ln \Gamma (-p^*), \label{e-act}
\end{eqnarray}
where
\begin{eqnarray}
p = - \frac{1 -2 \sigma}{2} + i \frac{m^2 + {\bf k}_{\perp}^2}{(m^2/ \epsilon)}.
\end{eqnarray}
Note the electromagnetic duality, $B = - iE$, holds for the unrenormalized QED action and thereby for the renormalized one, and that the unrenormalized action (\ref{e-act}) has the same UV divergent terms (\ref{ren}) with $\beta = i \epsilon$. By repeating the procedure for the case of magnetic field and using Eq. (\ref{log gam1}), we find the renormalized
action
\begin{eqnarray}
{\cal L}^{(1)} (\epsilon)= \frac{m^4}{(2 \pi)^2 (2 \epsilon)} \sum_{k =2}^{\infty} \frac{(-1)^k B_{2k}}{(2k)(2k-1)(2k-2)} \Bigl(\frac{1}{\epsilon} \Bigr)^{2k-1}.
\label{sp ser-E}
\end{eqnarray}
Note that the action (\ref{sp ser-E}) can be obtained by expanding the proper-time integral (\ref{prop in}).

Second, using Eq. (\ref{log gam2}), we get the renormalized action
\begin{eqnarray}
{\cal L}^{(1)} (\epsilon) = - \Bigl( \frac{m^2}{4 \pi \epsilon} \Bigr)^2 \Biggl[ \sum_{k =2}^{\infty}  \frac{\zeta (k, 1+ i \epsilon)}{(k+1)(k+2)} + \frac{1}{6} (\ln (i\epsilon) - \psi(1+ i\epsilon)) \Biggr].
\label{sp e-con}
\end{eqnarray}
Finally, summing over the Hurwitz zeta functions, we obtain the complex action for spinor QED
\begin{eqnarray}
{\cal L}^{(1)} (\epsilon)= - \Bigl( \frac{m^2}{4 \pi \epsilon} \Bigr)^2 \Bigl(2 \zeta'(-1, i \epsilon) +
(\epsilon^2 + i\epsilon  - \frac{1}{6}) \ln (i \epsilon) - \frac{\epsilon^2}{2} - \frac{1}{6} \Bigr).
\label{zeta E-act}
\end{eqnarray}
Then, the QED action (\ref{zeta E-act}) leads both to the vacuum polarization
\begin{eqnarray}
{\rm Re} {\cal L}^{(1)} (\epsilon) = - \Bigl( \frac{m^2}{4 \pi \epsilon} \Bigr)^2 \Bigl( \zeta'(-1, i \epsilon) +\zeta'(-1, - i \epsilon)
+ (\epsilon^2  - \frac{1}{6}) \ln (\epsilon) - \frac{\pi}{2} \epsilon - \frac{\epsilon^2}{2} - \frac{1}{6} \Bigr).
\label{E-vac pol}
\end{eqnarray}
and to the vacuum persistence
\begin{eqnarray}
{\rm Im} {\cal L}^{(1)} (\epsilon)=  \Bigl( \frac{m^2}{4 \pi \epsilon} \Bigr)^2 \Bigl[i \zeta'(-1, i \epsilon) - i \zeta'(-1, - i \epsilon)
+ \frac{\pi}{2}(\frac{1}{6} - \epsilon^2) - \epsilon \ln (\epsilon) \Bigr].
\label{E-vac per}
\end{eqnarray}
With the expansion formula (20) of Refs. \cite{adesizerbini02,fucci11}, the imaginary part equals to the vacuum persistence from the Schwinger action
\begin{eqnarray}
{\rm Im} {\cal L}^{(1)} (\epsilon)= \frac{1}{(2 \pi)} \Bigl( \frac{m^2}{4 \pi \epsilon} \Bigr)^2
\sum_{n =1}^{\infty} \frac{e^{-2 \pi \epsilon n}}{n^2}. \label{sch pair}
\end{eqnarray}
An interesting result of our method is that the Hurwitz zeta function in QED action (\ref{E-vac pol}) may be directly reconstructed from the imaginary part (\ref{sch pair}). Expanding the exponential functions in Eq. (\ref{sch pair}), summing over $n$ first in terms of the zeta functions, and using the reflection formula of the zeta function \cite{reflection}, we can show that
\begin{eqnarray}
\sum_{k =2}^{\infty} \frac{(- 2 \pi \epsilon)^k}{k!} \zeta (2-k) &=& -2 \epsilon \sum_{l =1}^{\infty} \Bigl(\frac{1}{2l} - \frac{1}{2l+1} \Bigr) \zeta (2l) (i \epsilon)^{2l} \nonumber\\
&=& \frac{1}{i} \Bigl(\zeta' (-1, 1- i \epsilon) - \zeta' (-1, 1+ i \epsilon) \Bigr) + \epsilon
\Bigl(\zeta' (0,1) - \zeta (0,1) \Bigr). \label{resum}
\end{eqnarray}
Thus, the formula in Ref. \cite{zeta-der} leads to the Hurwitz zeta function in the vacuum persistence (\ref{E-vac per}) and thereby the complex action (\ref{zeta E-act}) and the vacuum polarization (\ref{E-vac pol}). Note that the remaining terms, $\zeta (2)$, $\epsilon \zeta (1)$,
and $\epsilon^2 \zeta (0)$, have the connection with the other terms in Eq. (\ref{E-vac per}).

\section{Conclusion} \label{sec4}

The power series (\ref{sp ser}) of the Heisenberg-Euler-Schwinger action (\ref{prop in}) in the proper-time integral is asymptotically divergent and thus cannot be used for supercritical fields though the proper-time integral numerically converges for supercritical fields. However, any closed form for the proper-time integral has not been known yet. To overcome this methodological drawback, we have advanced a new method in the in-out formalism for computing QED actions suitable for supercritical magnetic and electric fields.
For that purpose, we have used the QED actions \cite{kly08,kly10,kim11} expressed by the Bogoliubov coefficients in the in-out formalism by Schwinger and DeWitt and employed the Ramanujan formula and the Whittaker-Watson formula for the gamma function. Then the QED action (\ref{zeta B-act}) in the magnetic field and the complex action (\ref{zeta E-act}) in the electric field are given by the Hurwitz zeta functions. The results of this paper are complementary to the QED actions in the proper-time integral from the same unrenormalized actions \cite{kly08,kly10,kim11}.

Our method is different from the zeta function method \cite{dittrichreuter} in that the unrenormalized QED action (\ref{qed ac}) in a magnetic field from the in-out formalism is given by a gamma function and the gamma function representation leads to the QED action (\ref{zeta B-act}) without summing over Landau levels and introducing a cutoff parameter for renormalization. A connection may exist between the zeta function regularization and the gamma function method in this paper, which cannot be clearly shown presently. More surprisingly, our method is still applicable to an electric field and leads to the complex QED action (\ref{zeta E-act}) in terms of the Hurwitz zeta function since the gamma function is well defined for a complex argument, though the zeta function regularization cannot be directly applied to unbounded motions of a charged particle in the electric field. The electromagnetic duality ($E = iB$) holds not only between the QED actions (\ref{zeta B-act}) and (\ref{zeta E-act}) but also between the unrenormalized actions.

\acknowledgments

The authors thank Christian Schubert and Yongsung Yoon for helpful discussions. S.P.K. would like to thank Don N. Page for the warm hospitality at
University of Alberta (UA), and Toshiki Tajima for the kind invitation as
a Norman Rostoker visiting professor at University of California, Irvine, and Rong-Gen Cai for the warm hospitality during the KIPTC Program ``Quantum Gravity, Black Holes, and String Theory'' at Kavli Institute for Theoretical Physics China (KITPC), Chinese Academy of Sciences (CAS), where this paper was completed.
H.K.L. would like to appreciate the warm hospitality at Kunsan National University. The visit by S.P.K. to UA and KITPC was supported in part by Basic Science Research Program through the National Research Foundation of Korea (NRF) funded by the Ministry of Education (NRF-2012R1A1B3002852).
This work was supported by the Research Center Program of IBS in Korea.


\begin{thebibliography}{99}
\bibitem{reflection} Formula 9.535 of Ref. \cite{gr-table}, which reads $2^{1-z} \Gamma (z) \zeta(z) \cos \Bigl( \frac{\pi}{2} z \Bigr) = \pi^z \zeta (1-z)$.
\end{thebibliography}
\end{document}